\documentclass[a4paper,11pt]{article}
\usepackage{pos}
\usepackage{siunitx}
\usepackage{graphicx}
\usepackage{multirow}

\title{A study on UV emission from clouds with Mini-EUSO}
 \ShortTitle{Mini-EUSO and clouds}

\author*[a,b]{Alessio Golzio}
\author[a,b]{Matteo Battisti}
\author[a]{Mario Bertaina}
\author[c]{Karl Bolmgren}
\author[d]{Giorgio Cambiè}
\author[e]{Marco Casolino}
\author[b]{Claudio Cassardo}
\author[f]{Roberto Cremonini}
\author[b]{Silvia Ferrarese}
\author[c]{Christer Fuglesang}
\author[b]{Massimiliano Manfrin}
\author[e]{Laura Marcelli}
\author[g]{Lech Piotrowski}
\author[h]{Kenji Shinozaki}

\affiliation[a]{INFN, Sezione di Torino,
  V. Giuria 1, Turin, Italy}

\affiliation[b]{University of Turin, Department of Physics,
  V. Giuria 1, Turin, Italy}

\affiliation[c]{KTH Royal Institute of Technology, Department of Physics,
  Brinellv\"agen 8, Stockholm, Sweden}

\affiliation[d]{University Tor Vergata, Department of Physics,
  V. della Ricerca Scientifica 1, Rome, Italy}

\affiliation[e]{INFN, Sezione di Roma Tor Vergata,
  V. della Ricerca Scientifica, Rome, Italy}

\affiliation[f]{ARPA Piemonte,
  V. Pio V 9, Turin, Italy}
  
\affiliation[g]{University of Warsaw,
  ul. Krakowskie Przedmieście 26/28, Poland}

\affiliation[h]{National Centre for Nuclear Research,
  Lodz, Poland}

\forColl{for the JEM-EUSO} % W/O "Collaboration"

\emailAdd{alessio.golzio@unito.it}

\abstract{Mini-EUSO is the first mission of the JEM-EUSO program located on the International Space Station. 
One of the main goals of the mission is to provide valuable scientific data in view of future large missions devoted to study Ultra-High Energy Cosmic Rays (UHECRs) from space by exploiting the fluorescence emission generated by Extensive Air Showers (EAS) developing in the atmosphere. 
A space mission like Mini-EUSO experiences continuous changes 
in atmospheric conditions, including the cloud presence. 
The influence of clouds on space-based observation is, therefore, an important topic to investigate as it might alter the instantaneous exposure for EAS detection or deteriorate the quality of the EAS images with consequences on the reconstructed EAS parameters. 
For this purpose, JEM-EUSO is planning to have an IR camera and a lidar as part of its Atmospheric Monitoring System. 
At the same time, it would be extremely beneficial if the UV camera itself would be able to detect the presence of clouds, at least in some specific conditions. 
For this reason, we analyze a few case studies by comparing the pixel count rates from Mini-EUSO during orbits with the cloud cover (as cloud fraction). 
This quantity is retrieved from the Global Forecast System (GFS) model at different height levels over the Mini-EUSO trajectory.
The results of this analysis are reported.}

\FullConference{37$^{\rm{th}}$ International Cosmic Ray Conference (ICRC 2021)\\\textsl{•}
		July 12th -- 23rd, 2021\\
		Online -- Berlin, Germany}

\begin{document}
\maketitle

\section{Introduction}
\begin{figure}
\center
\includegraphics[width=0.8\textwidth]{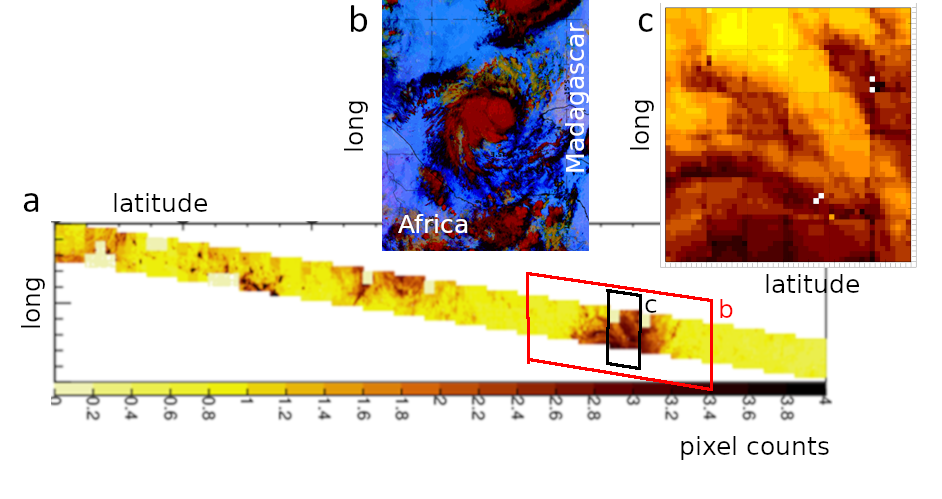}
\caption{a) composite image of the Mini-EUSO pixel counts over the Indian Ocean at approximately 18.38 UTC of December 5, 2019; b) a RGB Dust Composite image from the METEOSAT 8 meteorological satellite, showing the tropical cyclone seen by Mini-EUSO; c) a single image of the Mini-EUSO pixel matrix where the banded clouds of the cyclone are clearly visible. In the first part of the image, panel a), the dotted points are middle-low cumulus clouds.}
\label{fig:ciclone}
\end{figure}

The cloud cover determination is an open issue, and affects several experimental fields, from the physics of the atmosphere (meteorology and climatology), to astronomy, and to cosmic-ray detection system.
The Mini-EUSO telescope observes the Earth in the ultraviolet band (UV) \cite{Bacholle2021}, and is devoted to the study of ultra high energy cosmic rays coming from the space through their extensive air showers in the Earth's atmosphere.
From a first inspection of Mini-EUSO UV maps it was possible to state the capability of this telescope to detect tropospheric clouds (see Figure~\ref{fig:ciclone}).
In this particular event a tropical cyclone was detected in the Indian Ocean north-east of Madagascar Island (Figure~\ref{fig:ciclone}a, and a zoom in Figure~\ref{fig:ciclone}c).
The spiral-form of the cloud is also visible in the RGB Dust composite Image of METEOSAT8, the meteorological geostationary satellite, reported in Figure~\ref{fig:ciclone}b.

The Mini-EUSO telescope is on board of International Space Station (ISS) that completes an orbit around the Earth in $\approx 90$ minutes.
The covered area is a 350-km wide line oscillating between \ang{50} North-South.
The size of the considered area makes the cloud data retrieval from satellites very complex due to the fragmentation of the archives and the availability of images. 
So we decided, as a first step, to compare the UV maps with the cloud cover predicted by a Numeric Weather Prediction model (NWP) of the United States' National Weather Service, the Global Forecast System (GFS), that is available 24h-7days at a resolution of \ang{0.25} and with a 1-hour forecast step.

The actual NWPs compute the cloud fraction at each model level from pressure, cloud condensation mixing ratio, relative humidity and temperature data \cite{Xu1996}.
NWPs predict the place (grid point), the amount and the thickness of the cloud on a discrete vertical and horizontal grid.
The accuracy of NWPs is in general high (more than 70\%), with a forecast error less than 30\% in the first hours \cite{Ye2013}.
The cloud mask, resulting in a 2D, latitude-longitude, true/false fields, may be computed at high horizontal resolution using the NWPs cloud fraction data, following the same methodology applied to evaluate the cloud top height in \citep{Anzalone2019}.
In the present work, in order to obtain more accurate results, different thresholds are applied for high, medium and low clouds, and the cloud mask is compared with the UV maps from Mini-EUSO, to detect the cloud cover from UV data, a new application of this dataset.
The results are presented and discussed.

\section{Data and analysis}
Even though 41 Mini-EUSO sessions have been registered up to now, in this paper only 14 sessions are used, considering the availability of GFS high-resolution data (available only when the session was taken) and the UV maps. 
The available data for each of these sessions correspond to $\approx44$ minutes, coming from the first and/or the last part of the acquisition.

%\begin{table}[t]
%\center
%\caption{Mini-EUSO session data used and the number of points available in total or only over ocean and sea areas.}
%\label{tab1}
%\begin{tabular}{llccc}
%\hline\noalign{\smallskip}
%Session & Dates & All points & Sea points & Moon phase (\%)\\
%\hline\noalign{\smallskip}
%25 		& 2020-09-15 & 11674  	& 4968		& 12\\
%26		& 2020-09-25 & 9170		& 7763		& 55\\
%27		& 2020-12-09 & 9687		& 8038		& 34\\
%29		& 2021-01-05 & 10506	& 6647		& 71\\
%30		& 2021-01-09 & 10579	& 7542		& 27\\
%31		& 2021-01-12 & 10421 	& 7073		& 4\\
%32 		& 2021-01-16 & 10268	& 6944		& 6\\
%33		& 2021-01-21 & 9626		& 6502		& 46\\
%34		& 2021-02-01 & 9440		& 7946		& 84\\
%35		& 2021-02-03 & 9744		& 9310		& 75\\
%36		& 2021-02-06 & 10069	& 9948		& 41\\
%37		& 2021-02-11 & 10793	& 10394		& 2\\
%38  	& 2021-02-12 & 11085	& 10553		& 0\\
%40		& 2021-05-03 & 9385		& 5007		& 53\\
%41		& 2021-05-06 & 9242		& 4174		& 24\\
%\hline\noalign{\smallskip}
%\end{tabular}
%\end{table}

The analysis is conducted both on all Mini-EUSO available trajectories (over land and sea regions) and on the sea regions only, to avoid the contamination coming from antropogenic light.
The main light-source in space is due to the reflectance of the Moon (its phase is reported in Figure~\ref{fig:isto_hss} as percentage of full-Moon) and the nightglow/airglow signals, while the sources from the Earth are the cloud reflection (of space light), lightnings (not visible in the data because they were triggered by the sensor), boat lights or other atmospheric events.
The session considered span from 15 September 2020 to 6 May 2021.
The area compared in each session over sea span from \SI{2.61e6}{\square\kilo\metre} to \SI{6.60e6}{\square\kilo\metre}, while over sea and land span from \SI{5.73e6}{\square\kilo\metre} to \SI{7.30e6}{\square\kilo\metre}.

\section{Methods}
\subsection{UV maps and regridding}
Before the UV observations can be used, the possibility of inter-pixel differences (e.g. differences in pixel efficiency) must be accounted for. The different pixels of the Mini-EUSO detector are therefore subjected to a relative calibration using the lowest count of each pixel during the entire session as a baseline. This is possible because the physical environment (e.g. a cloudless ocean) producing this measurement should be the same for each pixel. 

The original pixel resolution of the Mini-EUSO telescope at ground is about \SI{6.3}{\kilo\metre} on the ground, and the matrix is composed by 48 by 48 pixels with a combined field of view of about $350 \times 350$~\si{\kilo\metre}. The photo-sensor registers the light in counts per pixel per time step (the acquisition is every \SI{41}{\milli\second}), and to make possible the comparison between the coarser GFS resolution a regridding operation is necessary.

This was done following three steps:
\begin{enumerate}
 	\item associate the center of each matrix-pixel-time observation with the corresponding latitude and longitude coordinates. 
 	This is done by projecting the shape of the detector pixel array onto the surface of the Earth by knowing the position of the ISS and the detector geometry. 
 	These calculations are made in polar coordinates, with the radial distance $\rho_{obs}$ between an observation on the ground and the ISS subpoint as 
 	\begin{equation} 
 		\rho_{obs}= h_{ISS}\tan(FoV_{p})
 	\end{equation} 
	where $rho_{p}$ is the radial distance to the pixel from the center of the detector, $h_{ISS}$ is the altitude of the ISS and $FoV_{p}$ is the part of the field of view taken up by one pixel. The polar coordinates are subsequently transformed geographic coordinates.  
 	\item collect overlapping pixels by making spatial averages (in \SI{41}{\milli\second} the ISS moves of \SI{\approx 300}{\metre})
 	\item regrid on a \ang{0.25} longitude-latitude grid at the same center-points of GFS, considering the average of the pixel-counts within each bin formed by the grid. 
\end{enumerate}

\subsection{Cloud mask computation}
\subsubsection{GFS cloud masks}
The GFS data\footnote{The GFS operative data are real-time retrived from \url{https://nomads.ncep.noaa.gov/cgi-bin/filter_gfs_0p25_1hr.pl}} used in the present work consists of the cloud fraction fields with a grid spacing of \ang{0.25} and 39 vertical levels (from \SI{1e5}{\pascal} to \SI{5e3}{\pascal}), resulting in 40491360 points for each session.

Cloud fraction values span from 0 to 1, where 0 is clear sky and 1 overcast sky.
In order to calculate the cloud mask for low, middle and high layers, the cloud fraction maximum is identified in the three layers and its value is compared with the following thresholds: Low clouds ($z < 2$ km a.s.l.) are present if maximum low-cloud fraction overpass $0.2$; Medium clouds ($2 < z < 6$ km a.s.l.) are present if maximum medium-cloud fraction overpass $0.4$; High clouds ($6 < z < 18$ km a.s.l.) are present if maximum high-cloud fraction overpass $0.8$.
This set of thresholds was choosen after studies \cite[e.g.][]{Anzalone2019} that shows the different ability of the global model to detected the three levels of clouds.
Three cloud cover masks are computed respectively for low, medium and high layers, the total cloud mask is the union of these three masks.

\subsubsection{UV cloud mask}
The UV cloud mask is computed on the UV mean counts re-sampled at the same grid spacing of GFS, using three different thresholds: 1, 5 and 10 counts per grid point.

\subsection{Contingency tables}
\label{ss_tables}
To assess the differences in the cloud masks a contingency table is built between each cloud level and UV threshold and for every session.
The cloud mask from the global model is considered the truth, while the cloud mask from Mini-EUSO UV band is considered the test, so the contingency table is built following Figure~\ref{fig1}.

\begin{figure}
\center
\includegraphics[width=0.45\textwidth]{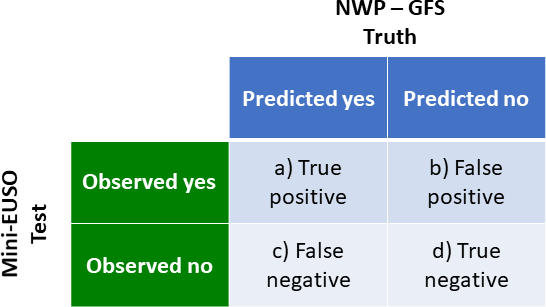}
\caption{The contingency table used in this study. Letters refer to the calculated indexes of \ref{ss_tables}}
\label{fig1}
\end{figure}

The two considered indexes are the Accuracy (Proportion Correct, PC) and the Heidke Skill Score \cite[HSS,][]{Heidke1926}, the first indicates the hits (true positive and true negative) with respect the overall cases, while the HSS indicates the quality of the forecast, where the PC measure is scaled with the reference value from correct forecasts due to chance. 

PC is defined as (refering to Figure \ref{fig1}):

\begin{equation}
PC = \frac{a+d}{a+b+c+d} = \frac{a+d}{n}
\end{equation}

The perfect value for PC is the unity (or here a percentage if PC is multiplied by 100). Its maximum-likelihood estimate is 
\begin{equation}
E = \bigg(\frac{a+c}{n}\bigg)\left(\frac{a+b}{n}\right) + \left(\frac{b+d}{n}\right)\left(\frac{c+d}{n}\right)
\end{equation}
HSS is defined as:
\begin{equation}
HSS = \frac{PC - E}{1-E} = \frac{2(ad-bc)}{(a+b)(b+d)+(a+c)(c+d)}
\end{equation}

The HSS ranges between $-\infty$ and 1, where 1 is the perfect forecast, 0 indicates no-skill and negative values indicates that the chance forecast is better.

\section{Results and discussion}\label{ss:genres}
The general statistics (Table~\ref{tab2}) show the higher performances with the UV threshold set at one count per pixel, with slightly better results in the accuracy of low clouds over the sea (61.0\%), while over the entire dataset, middle (total) cloud cover obtain the best value (60.8\%).
The HSS shows similar results regarding the UV thresholds but with a more significant gap.
Again the low-clouds prediction obtained the best results: $0.12$ on the sea and $0.17$ over land and sea together.

\begin{table}[h]
\caption{Averaged results over the 14 sessions considered. Medium-cloud values represent also the Total cloud cover statistics.}
\label{tab2}
\center
\begin{tabular}{p {3 cm} crrrrrr}
\hline\noalign{\smallskip}
Points location & UV threshold & \multicolumn{3}{c}{Accuracy ($100 PC$)} & \multicolumn{3}{c}{HSS}\\
  & & High & Medium & Low & High & Medium & Low \\
\hline\noalign{\smallskip}
\multirow{3}{*}{Sea (\SI{6.48e7}{\square\kilo\metre})} & 1 & 48.3 & 59.6 & 61.1 & 0.019 & 0.072 & 0.119 \\
 						      & 5 & 49.0 & 40.1 & 57.0 & 0.009 & 0.047 & 0.088 \\
 						      & 10& 50.5 & 36.3 & 56.3 & 0.015 & 0.047 & 0.076 \\
\multirow{3}{*}{All (\SI{9.48e7}{\square\kilo\metre})} & 1 & 47.4 & 60.8 & 56.7 & 0.050 & 0.108 & 0.167 \\
							  & 5 & 48.8 & 42.2 & 58.3 & 0.009 & 0.052 & 0.091 \\
							  & 10& 50.5 & 38.1 & 58.0 & 0.015 & 0.048 & 0.080 \\
\hline\noalign{\smallskip}
\end{tabular}
\end{table}

\subsection{Sea-data results}\label{ss:seares}
We considered only sea points to improve the analysis and remove the pixel counts related to land lights.
The first question was to find a possible correlation with the Moon phase, as it is the most important light source during the night. 
Linear regression Pearson correlation coefficient is calculated between the accuracy and the HSS indexes and the Moon phase.
The considered cases are the three cloud levels with the three UV thresholds on the two indexes (Accuracy and HSS).
The correlation shows low values mainly due to the limited statistics, never higher than $0.17$, and the most correlated series with the moon phase is the low cloud HSS with the higher UV threshold ($R^2 = 0.166$).
The poor correlation values and the flatness of linear regression lines indicate that the moon does not affect the detection of clouds.
Moreover, considering some single situation, the light curve from Mini-EUSO suggests that the most appropriate threshold is 1 (as also deducted from beginning of Section~\ref{ss:genres} and Table~\ref{tab2}).

\begin{figure}
\center
\includegraphics[width=0.8\textwidth]{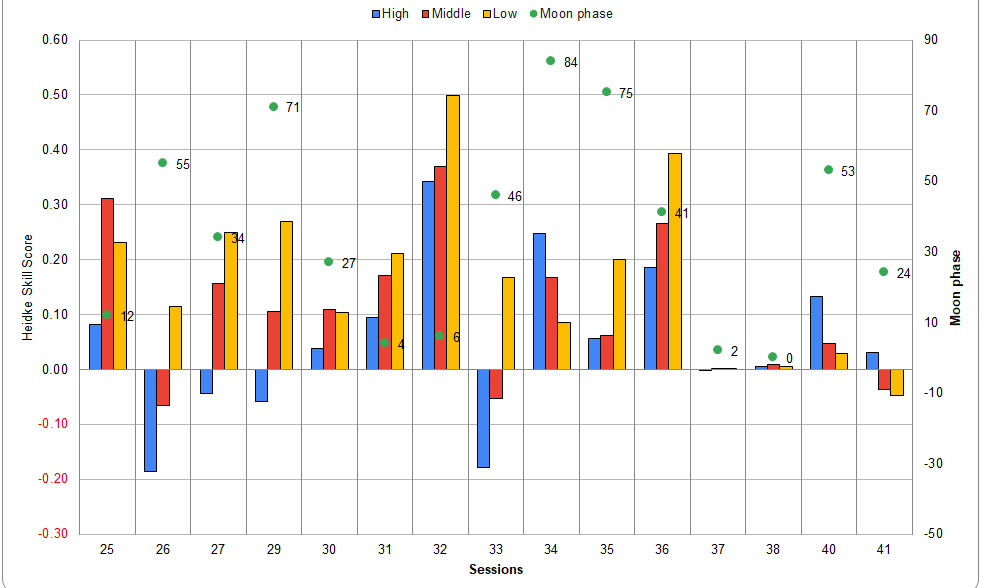}
\caption{The histogram of HSS with the UV threshold 1 and the sea data only.}
\label{fig:isto_hss}
\end{figure}

\begin{figure}
\center
\includegraphics[width=0.8\textwidth]{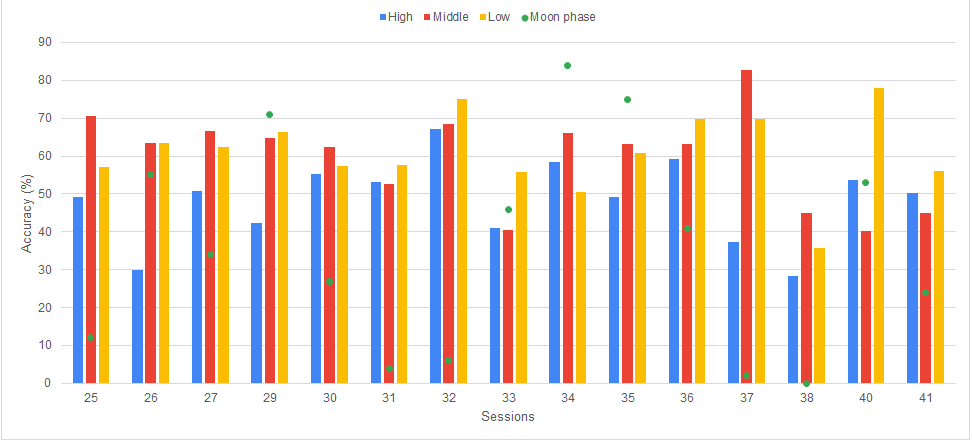}
\caption{The histogram of Accuracy with the UV threshold 1 and the sea data only.}
\label{fig:isto_acc}
\end{figure}

The accuracy values of UV threshold 1 are between 30\% and 82\%, but sometimes very high accuracy values present very low (near zero) HSS, indicating a poor agreement between observation and prediction. 
This is the case of session 37 (Figures~\ref{fig:isto_hss} and \ref{fig:isto_acc}).

Analyzing the two indexes together, two sessions appear especially interesting: 32 and 36. 
In these two cases, the Moon phase was respectively 6\% and 41\%, confirming the slight influence of the quantity of moon light, and  accuracy is 70.14\% and 64.06\%, while HSS is 0.40 and 0.28. 
In particular HSS have very high and promising values.

\begin{figure}[t]
\center
\includegraphics[width=0.75\textwidth]{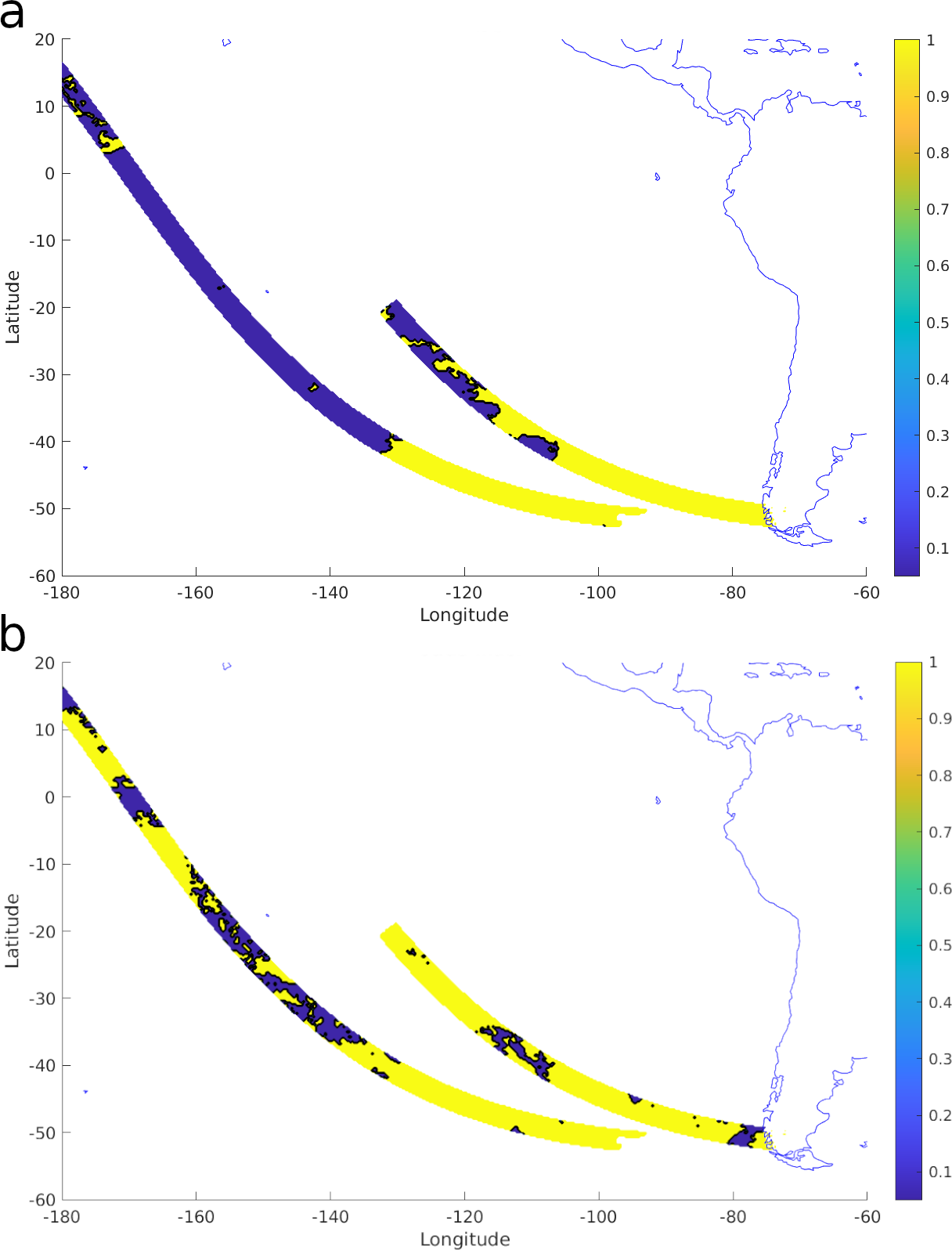}
\caption{Session 36, 6 February 2021. a) The cloud mask computed on the UV counts from Mini-EUSO, the visible points are the pixels that pass the threshold of 1 count per pixel; b) the cloud mask from the GFS cloud fraction, as the union of low, medium and high clouds. In both cloud masks 1 represents overcast sky and 0 clear sky.}
\label{fig:cloud}
\end{figure}

In Figure~\ref{fig:cloud} the cloud masks calculated for session 36 is shown. It is possible to see that when the threshold of 1 is overpassed on pixel counts clouds are present, while in clear sky zones the UV map shows no pixel counts (white is below the threshold).

\section{Conclusions}
This study, considering the complexity of cloud detection and forecast, shows the ability of Mini-EUSO UV sensor to detect cloud cover during its operative time. The best UV threshold to compute the cloud mask appears to be one count per pixel, and the overall cloud detection was good (mainly in the two reported examples of session 32 and 36), also considering the degradation of the UV maps to reach the same grid spacing of GFS.
From the first inspection showed in Figure~\ref{fig:ciclone}, and from the complete statistics over the 14 session considered, it is possible to state that the Mini-EUSO sensor is able to capture the middle and low cloud better than high clouds, that usually are thinner and have a smaller optical depth. 
However they are able able to modify the detection of the EAS.

Future improvements and investigations will increase the reference NWP resolution and reach a greater detail in cloud cover to be compared.

\acknowledgments
This work was supported by State Space Corporation ROSCOSMOS, by the Italian Space Agency through the ASI INFN agreement n. 2020-26-HH.0 and contract n. 2016-1-U.0, by the French space agency CNES, National Science Centre in Poland grant 2017/27/B/ST9/02162.

This research has been supported by the Interdisciplinary Scientific and Educational School of Moscow University "Fundamental and Applied Space Research". 
The article has been prepared based on research materials carried out in the space experiment "UV atmosphere".

\bibliographystyle{JHEP}   
\bibliography{bib_ICRC2021.bib} 

\providecommand{\href}[2]{#2}\begingroup\raggedright\begin{thebibliography}{1}

\bibitem{Bacholle2021}
S.~Bacholle, P.~Barrillon, M.~Battisti, A.~Belov, M.~Bertaina, F.~Bisconti
  et~al., \emph{Mini-{E}{U}{S}{O} mission to study earth {U}{V} emissions on
  board the {I}{S}{S}},
  \href{https://doi.org/10.3847/1538-4365/abd93d}{\emph{The Astrophysical
  Journal Supplement Series} {\bfseries 253} (2021) }.

\bibitem{Xu1996}
K.M.~Xu and D.A.~Randall, \emph{A semiempirical cloudiness parameterization for
  use in climate models},
  \href{https://doi.org/10.1175/1520-0469(1996)053<3084:ASCPFU>2.0.CO;2}{\emph{Journal
  of Atmospheric Sciences} {\bfseries 53} (1996) 3084}.

\bibitem{Ye2013}
Q.-Z.~Ye and S.-S.~Chen, \emph{The ultimate meteorological question from
  observational astronomers:how good is the cloud cover forecast?},
  {\emph{Monthly Notices of the Royal Astronomical Society} {\bfseries 428}
  (2013) 3288}.

\bibitem{Anzalone2019}
A.~Anzalone, M.E.~Bertaina, S.~Briz, C.~Cassardo, R.~Cremonini, A.J.~de~Castro
  et~al., \emph{Methods to retrieve the cloud-top height in the frame of the
  {J}{E}{M}-{E}{U}{S}{O} mission}, {\emph{IEE Transctions on geoscience and
  remote sensing} {\bfseries 57} (2019) 304}.

\bibitem{Heidke1926}
P.~Heidke, \emph{Berechnung des erfolges und der g\"{u}te der
  windst\"{a}rkevorhersagen im sturmwarnungsdienst},
  \href{https://doi.org/10.1080/20014422.1926.11881138}{\emph{Geografiska
  Annaler} {\bfseries 8} (1926) 301}.

\end{thebibliography}\endgroup
%\begin{thebibliography}{99}
%\bibitem{Xu-Randall1996}
%....

%\end{thebibliography}

%% Full authors list (ONLY FOR COLLABORATIONS)
%\clearpage
\noindent
{\Large\bf Full Authors List: The JEM-EUSO Collaboration\\}

\vspace*{0.5cm}

\begin{sloppypar}
{\small \noindent 
G.~Abdellaoui$^{ah}$, 
S.~Abe$^{fq}$, 
J.H.~Adams Jr.$^{pd}$, 
D.~Allard$^{cb}$, 
G.~Alonso$^{md}$, 
L.~Anchordoqui$^{pe}$,
A.~Anzalone$^{eh,ed}$, 
E.~Arnone$^{ek,el}$,
K.~Asano$^{fe}$,
R.~Attallah$^{ac}$
% ...
}
\end{sloppypar}
\vspace*{.3cm}

%%\newpage
{ \footnotesize
\noindent
% Algeria (Dezember 2013) - 7 institutes
% ...
$^{ac}$ LPR at Dept. of Physics, Faculty of Sciences, University Badji Mokhtar, Annaba, Algeria \\
% ...
$^{ah}$ Telecom Lab., Faculty of Technology, University Abou Bekr Belkaid, Tlemcen, Algeria \\
% Bulgaria ready (02042012)  - 1 institutes 
% ...
$^{cb}$ Universit\'e de Paris, CNRS, AstroParticule et Cosmologie, F-75013 Paris, France\\
% ...
$^{ed}$ Istituto Nazionale di Fisica Nucleare - Sezione di Catania, Italy\\
% ...
$^{eh}$ INAF - Istituto di Astrofisica Spaziale e Fisica Cosmica di Palermo, Italy\\
% ...
$^{ek}$ Istituto Nazionale di Fisica Nucleare - Sezione di Torino, Italy\\
$^{el}$ Dipartimento di Fisica, Universita' di Torino, Italy\\
% ...
$^{fe}$ Institute for Cosmic Ray Research, University of Tokyo, Kashiwa, Japan\\ 
% ...
$^{fq}$ Nihon University Chiyoda, Tokyo, Japan\\ 
% ...
$^{md}$ Universidad Polit\'ecnia de madrid (UPM), Madrid, Spain\\
% ...
$^{pd}$ University of Alabama in Huntsville, Huntsville, AL; USA\\
$^{pe}$ Lehman College, City University of New York (CUNY), NY, USA\\
% ...
%16 Leerzeilen in Affils.
}

%\noindent
%{\scriptsize \bf In Summary: \hspace{0.4cm} 293 authors \hspace{0.4cm} 85 institutes \hspace{0.4cm} 17 countries}

%\noindent
%{\scriptsize out in December 2019, therefore still author until end of 2020: Napolitano, Franceschi, Contino, Shirahama, Turriziani, Tone,  Kawasaki, Hachisu}

\vspace*{0.5cm}

%\section*{Full Authors List: \Coll\ Collaboration}
%
%\noindent \textbf{Note comment afterwards:} Collaborations have the possibility to provide an authors list in xml format which will be used while generating the DOI entries making the full authors list searchable in databases like Inspire HEP. For instructions please go to icrc2021.desy.de/proceedings or contact us under icrc2021proc@desy.de.\\
%
%\scriptsize
%\noindent
%first.author$^1$, 
%second.author$^2$, 
%third.author$^3$ % .... more names
%and 
%last.author$^{n}$ \\
%
%\noindent
%$^1$first.affiliation.
%$^2$second.affiliation. % .... more affiliation
%$^{m}$last.affiliation.

\end{document}